\documentclass[a4paper,fleqn,usenatbib]{mnras}

\usepackage{newtxtext,newtxmath}

\usepackage[T1]{fontenc}
\usepackage{ae,aecompl}
\usepackage{longtable}

\newcommand{\ms}{$\,$M$_\mathrm{\odot}$}

\newcommand{\be}{\begin{equation}}
\newcommand{\ee}{\end{equation}}

\newcommand{\el}[2]{\ensuremath{^{#1}\mathrm{#2}}}

\def \mnras {MNRAS}
\def \araa {Annual Reviews of Astronomy \& Astrophysics}
\def \aap {A\&A}
\def \apj{ApJ}
\def \apjl{ApJL}

\def \pasa {Publications of the Astronomical Society of Australia}

\usepackage{graphicx}	
\usepackage{amsmath}	
\usepackage{longtable}

\title[Barium giants and mass accretion]{The formation of barium giants via mass accretion in binary systems}

\author[R.~J. Stancliffe]{
Richard J. Stancliffe$^{1}$\thanks{E-mail: R.Stancliffe@hull.ac.uk}
\\
$^{1}$E.~A. Milne Centre for Astrophysics, Department of Physics \& Mathematics, University of Hull, Cottingham Road, Kingston upon Hull, HU6 7RX\\
}

\date{Accepted XXX. Received YYY; in original form ZZZ}

\pubyear{2020}

\begin{document}
\label{firstpage}
\pagerange{\pageref{firstpage}--\pageref{lastpage}}
\maketitle

\begin{abstract}
We examine the composition of barium stars in the context of mass transfer from an asymptotic giant branch (AGB) companion. We accrete between 0.01 and 0.5\ms\ of AGB ejecta on to low mass companions of [Fe/H] = -0.25 at the ages expected for the end of the lives of AGB stars of 2.5, 3 and 4\ms. In each case, we form a star of 2.5\ms\ which is thought to be a typical barium star mass. We discuss the extent of dilution of accreted material as the star evolves, and describe the impact on the surface abundances. For accretion from a 2.5\ms\ primary, if the secondary's initial mass is 2.45\ms\ or more, accretion takes place when the secondary is undergoing core helium burning. Using data from the sample of De Castro et al., we attempt to fit the observed properties of 74 barium giants using the models we have computed. We find that all but six of these objects are best fit using ejecta from 2.5\ms\ (32 objects) or 3\ms\  (36 objects) AGB stars. Higher accretion masses are typically required when accreting from a lower mass companion. We find accretion masses that are broadly consistent with recent hydrodynamical simulations of wind mass transfer, though the accretion efficiency is toward the upper limit found in these simulations. For the 18 stars with reported orbital periods, we find no strong correlations between period and accretion mass.
\end{abstract}




\section{Introduction}

Barium stars are a class of stellar object that were first identified through the anomalous intensity of their CH bands, and strong lines of singly-ionised strontium and barium \citep{1951ApJ...114..473B}. The study of elemental abundances in the barium star HD~46407 led \citet{1957ApJ...126..357B} to conclude that the heavy element pattern was produced via the slow neutron capture process \citep{1957RvMP...29..547B}. 
\citet{1980ApJ...238L..35M} were the first to establish the binary nature of Ba stars. Out of a sample of 11 stars of the Ba2-Ba5 classes \citep[which were defined by][using the strength of barium lines]{1965MNRAS.129..263W}, significant radial velocity variations were found for 9 of them. Two of the stars in this sample had been studied for a complete, or nearly complete, orbital period which allowed the authors to conclude that the unseen companion was of low mass. The presence of radial velocity variations, together with enhanced levels of carbon and s-process elements suggests that barium stars are formed in binary systems, where an asymptotic giant branch (AGB) star is able to transfer material to a lower mass companion \citep{1980ApJ...238L..35M}. This same scenario is also invoked to explain the origins of other classes of stars: the CH stars, Tc-poor S stars and the carbon-enhanced metal-poor stars of the CEMP-s subclass.

Barium stars are useful objects for study because they provide information on both low-mass star nucleosynthesis and the physics of binary systems. As objects that have been contaminated with the ejecta of asymptotic giant branch stars, they have been used to probe the production of s-process nuclei \citep[e.g.][]{2001ApJ...557..802B,2018A&A...620A.146C}. The orbital properties of barium stars can help to constrain the physics of mass transfer \citep[e.g.][]{2008A&A...480..797B,2010A&A...523A..10I}, and tidal interactions \citep[e.g.][]{2020arXiv200505391E}. 

From a theoretical perspective, barium stars should be found in both unevolved and evolved stages. Assuming both primary and secondary star are not too close in initial mass, the primary star will transfer matter to the secondary star while the latter is on the main sequence. This event marks the birth of the barium star. The secondary will itself eventually evolve, leaving the main sequence and becoming a giant. Assuming it was polluted with a sufficient quantity of material from its primary, the secondary will remain a barium star for the rest of its existence. Observationally, barium stars are divided into Ba giants and dwarf barium stars. In recent years, both types of barium star have become increasingly well studied, both in terms of their orbital properties \citep{2019A&A...626A.128E, 2019A&A...626A.127J} and chemical abundances \citep{2016MNRAS.459.4299D}.  In addition to the increased sample sizes now available to work with, data from the Tycho-Gaia Astrometric Solution \citep[TGAS,][]{2015A&A...574A.115M,2016A&A...595A...4L} has also helped to constrain the masses of the stars involved. \citet{2017A&A...608A.100E} were able to obtain the masses of over 400 barium stars, both giants and dwarfs. For the giants they obtain a distribution of masses which peaks at 2.5\ms\, with a broad tail running up to about 4.5\ms. This distribution is based upon an assumed metallicity of [Fe/H]$=-0.25$, though subsequent work by \citet{2019A&A...626A.127J} using appropriate metallicities obtained a similar mass distribution.

In this paper, we make use of the elemental abundances derived by \citet{2016MNRAS.459.4299D}, and in combination with the information on the mass distribution of barium giants derived by \citet{2017A&A...608A.100E}, we attempt to model the properties of these stars. We compute the full evolution of future barium stars from the birth of the unpolluted secondary on the main sequence, through the accretion of material from their AGB companions, up to the point that they become AGB stars themselves. Such models give insight into the nature of the mass of the companion AGB star, and how much material it is able to transfer.

\section{The {\sc{STARS}} code}

The stellar models for this work were all produced using the Cambridge stellar evolution code, {\sc stars} \citep{1971MNRAS.151..351E, 1995MNRAS.274..964P, 2009MNRAS.396.1699S}, which simultaneously solves the  equations of stellar structure and the chemical evolution of seven species: \el{1}{H}, \el{3}{He}, \el{4}{He}, \el{12}{C}, \el{14}{N}, \el{16}{O} and \el{20}{Ne}. A further 40 isotopes which are energetically unimportant are followed via a post-processing step, calculated immediately after the convergence of each timestep of the evolution model \citep{2005MNRAS.360..375S}. Models are evolved from the pre-main sequence phase with 499 meshpoints, and no mass loss was included. A mixing length of $\alpha=2.025$ is employed, based on the calibration of \citet{2016A&A...586A.119S} which was performed using a Solar model with the abundances of \citet{2009ARA&A..47..481A}. Convective overshooting is included in the model using the prescription of \citet{1997MNRAS.285..696S}, with $\delta_\mathrm{ov}=0.15$ based on calibration to eclipsing binary stars \citep{2015A&A...575A.117S}.

\begin{figure}
\begin{center}
\includegraphics[width=\columnwidth]{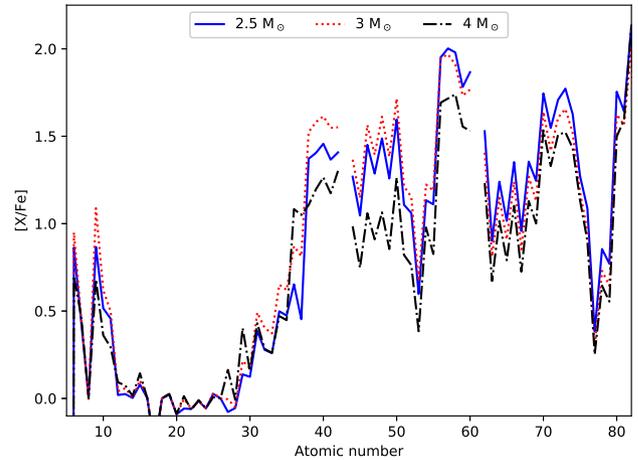}
\end{center}
\caption{[X/Fe] as a function of atomic number for the three \citet{2016ApJ...825...26K} AGB models used. These abundances are the average of all material ejected over the star's lifetime.}
\label{fig:Yields}
\end{figure}

The method used for constructing the models is the same as employed by \citet{2007A&A...464L..57S} and \citet{2008MNRAS.389.1828S} for the formation of carbon-enhanced, metal-poor stars. A model is evolved up to a certain age which is determined by the lifetime of its hypothetical AGB companion. At this point, material of the average composition of the ejecta of the AGB star is accreted onto the star's surface at a rate of $10^{-6}$\ms\ per year, until the target mass is reached. Models are constructed by accreting 0.01, 0.025, 0.05, 0.1, 0.25 or 0.5\ms\ of material onto a star until a mass of 2.5\ms\ is reached, which we take to be the typical mass of a barium star based on the work of \citet{2017A&A...608A.100E}. We also take [Fe/H] = -0.25 as the typical metallicity of a barium star based on the same work. We consider three masses of AGB star: 2.5, 3 and 4\ms. The time at which accretion takes place is computed from our own models, but we take the composition of the ejecta from \citet{2016ApJ...825...26K}. Some details of these models are given in Table~\ref{tab:properties}, and the abundance patterns of the three models as a function of atomic mass are shown in Fig.~\ref{fig:Yields}.

Because barium abundances are rarely determined for barium stars, it is common to describe the overall s-process enrichment by an average of some of the other elements whose abundances are more readily determined. \citet{2016MNRAS.459.4299D} define [s/Fe] as the mean of five abundances: [Y/Fe], [Zr/Fe], [La/Fe], [Ce/Fe], and [Nd/Fe]. We give the values of these elements in Table~\ref{tab:properties}, along with the resulting [s/Fe] value. \citet{2016MNRAS.459.4299D} give [s/Fe] $>+0.25$ as the definition of a barium star, based on high resolution spectra and for simplicity we adopt this same definition. A more detailed discussion of barium star classifications can be found in \citet{2019A&A...626A.127J}, and these authors advocate that 1 dex for [La/Fe] and [Ce/Fe] represents the transition between mild and strong barium stars, while no mild bariums stars exist for [Ce/Fe]$<0.2$.

\begin{table*}
\begin{center}
\begin{tabular}{ccccccccccc}
Initial Mass & Lifetime & $\log \epsilon$(Li) & \el{12}{C} & \el{14}{N} & [Y/Fe] & [Zr/Fe] & [La/Fe] & [Ce/Fe] & [Nd/Fe] & [s/Fe] \\
(\ms) & (Myrs) \\
\hline
2.5 & 677.0 & 2.35 & $1.11\times10^{-2}$ & $2.17\times10^{-3}$ & 1.41 & 1.46 & 2.00& 1.98 & 1.87 & 1.74 \\
3.0 & 406.0 & 1.49 & $9.10\times10^{-3}$ & $2.75\times10^{-3}$ & 1.58 & 1.61 & 1.97 & 1.91 &1.77 & 1.77 \\
4.0 & 191.1 & 3.71 & $6.03\times10^{-3}$ & $1.08\times10^{-3}$ & 1.20 & 1.26 &  1.72 & 1.74 & 1.53 & 1.49  \\
\hline
\end{tabular}
\end{center}
\label{tab:properties}
\caption{Total lifetimes and composition for select species of the AGB models used. The abundances of \el{12}{C} and \el{14}{N} are given as mass fractions. These abundances are the average of all material ejected over the star's lifetime.}
\end{table*}

\section{Results}
\label{sec:models}

\begin{figure}
\includegraphics[width=\columnwidth]{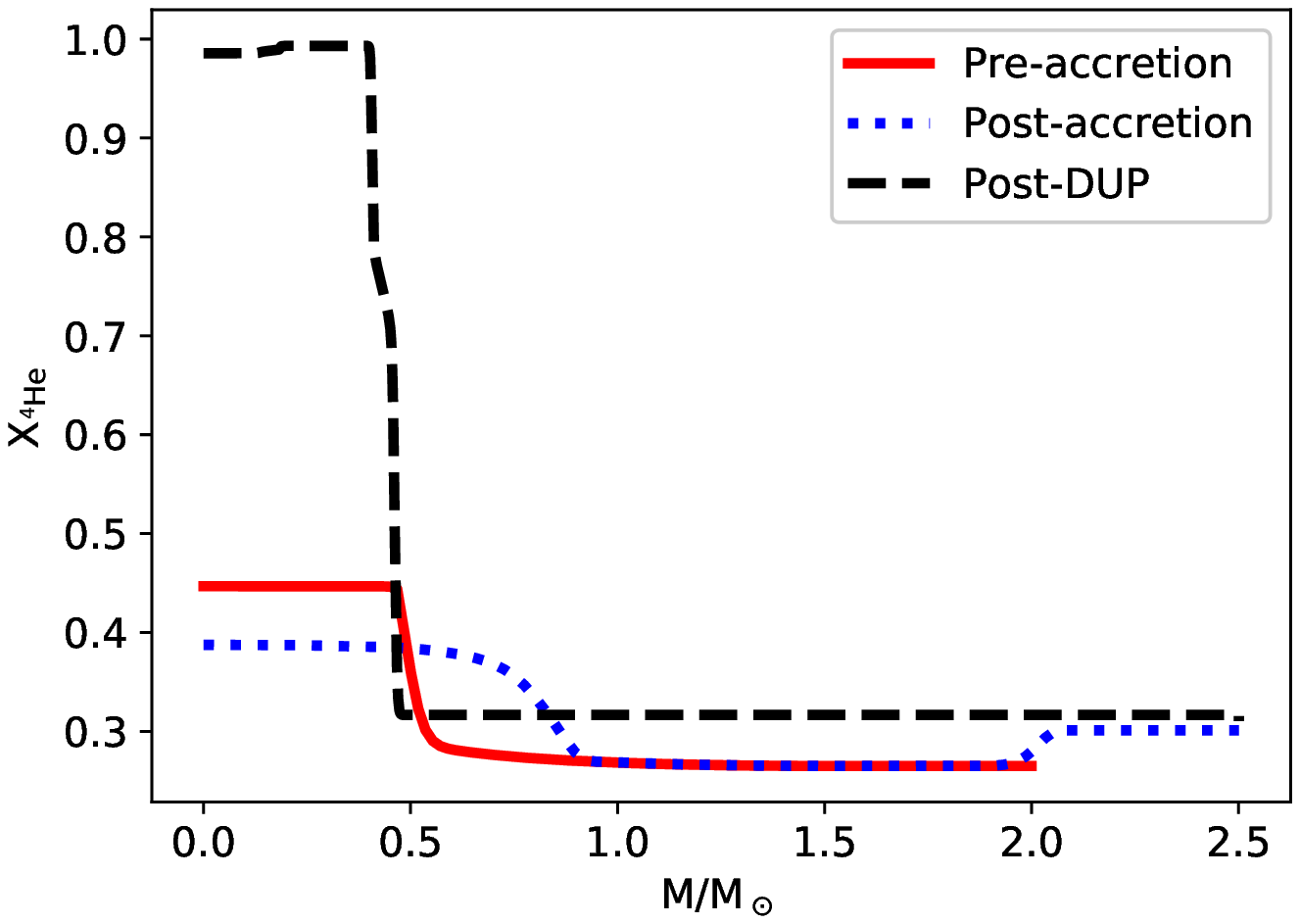}
\includegraphics[width=\columnwidth]{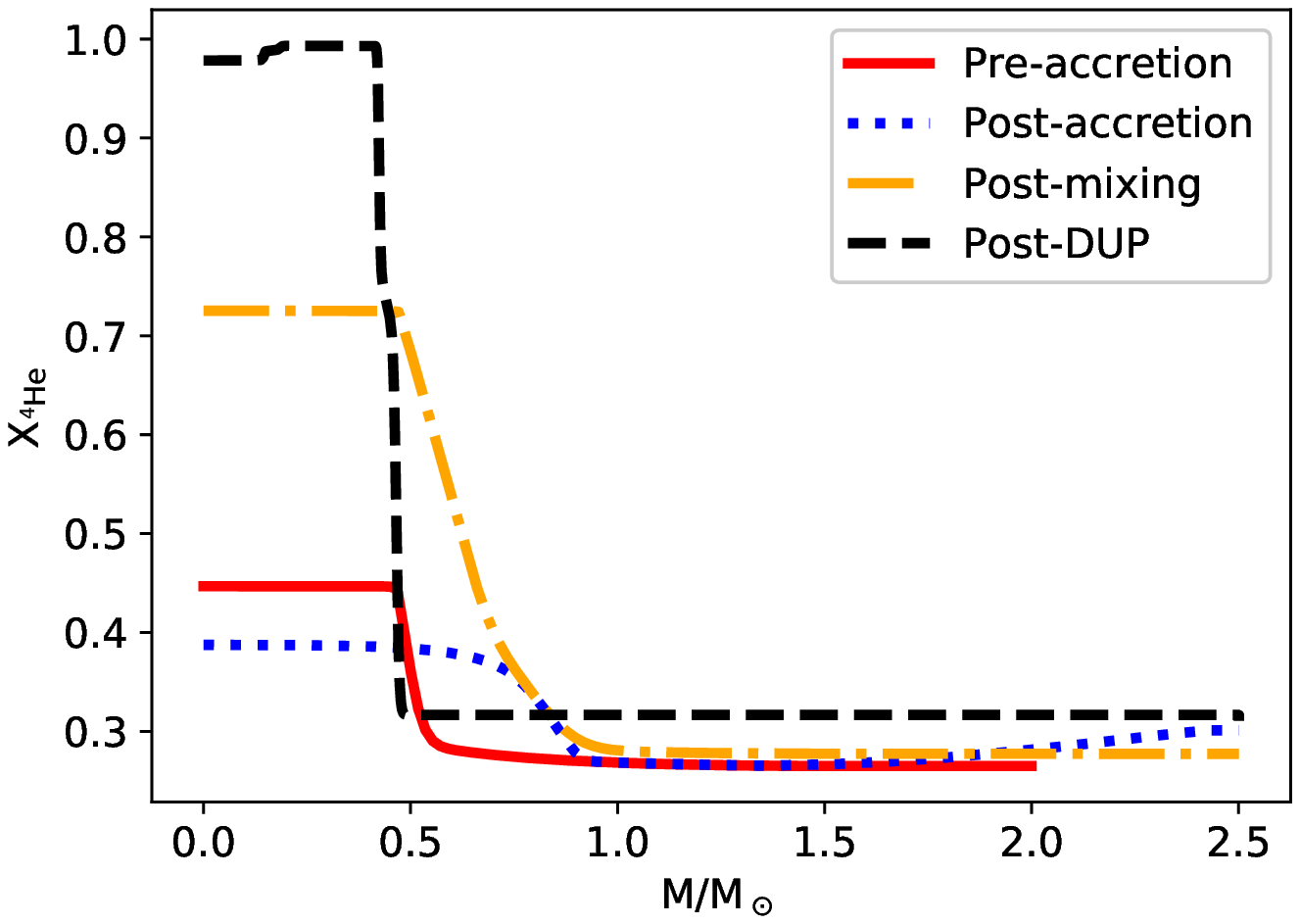}
\caption{Helium abundance as a function of mass coordinate for the 2\ms\ star, accreting 0.5\ms\ of material from a 3\ms\ companion. {\bf Upper panel:} Model including only convective mixing. {\bf Lower panel:} Model including both convective and thermohaline mixing.}
\label{fig:m2M3}
\end{figure}

To describe the basic features of these models, we discuss the case of accreting 0.5\ms\ of material on to a companion of 2\ms. We begin with the case of accreting from a 3\ms\ AGB star, which starts to transfer mass at around 400 Myrs. At this point, the 2\ms\ star has reached a core helium abundance of around 0.44, and its convective core contains around 0.5\ms\ of material. As material is accreted, the convective core expands to reach a mass co-ordinate of around 0.75\ms, ingesting fresh hydrogen and rejuvenating the star. The core helium abundance is reduced to below 0.4 (see Fig.~\ref{fig:m2M3}). If we only account for mixing by convection, the accreted layer remains at the star's surface until a convective envelope develops as the star ascends the giant branch. At its maximum depth, the convective envelope reaches down to a mass coordinate of around 0.5\ms\ and penetrates the regions of the core that became mixed during the accretion event. This point represents the maximum dilution of the accreted material as the convective envelope never again reaches this depth in the remainder of the star's life. If we define the dilution factor, $d$, as:
\begin{equation}
d = {M_\mathrm{acc} \over M_\mathrm{mix}}
\end{equation}
where $M_\mathrm{acc}$ is the mass of accreted material, and $M_\mathrm{mix}$ is the mass over which it is mixed, then $d$ = 0.247 at this point. The maximum depth of FDU is insensitive to the small variations in composition encountered in this work, and so the post-FDU dilution factor does not vary when we accreted the same amount of mass from different companion stars.

Because the accreted material has a higher mean molecular weight that the envelope of the recipient star, the accreted material should not remain on the surface. Thermohaline convection, should act to mix the material into the interior \citep{1972ApJ...172..165U,1980A&A....91..175K}. Thermohaline (TH) mixing is rapid compared to the main sequence lifetime and so the accreted material is rapidly mixed in. The depth that thermohaline mixing extends to is set by the point at which the mean molecular weight in the layers below exceeds that in those above. This condition is met at the boundary of the convective core (see the dot-dashed line in the lower panel of Fig.~\ref{fig:m2M3}). TH mixing homogenises the envelope and once equilibrium is reached the dilution factor is 0.312. As the star ascends the giant branch, the convective envelope reaches deeper than the maximum depth of TH mixing, resulting in a drop in the dilution factor (to 0.247) as the star reaches the upper regions of the giant branch. Note this behaviour is different to the case of the low-mass CEMP stars modelled by \citet{2008MNRAS.389.1828S}, where TH mixing often extends beyond the depth reached by the convective envelope at maximum penetration. This makes giant Ba stars far less sensitive to mixing physics than CEMP stars.

When accreting material from a 2.5\ms\ companion, the star no longer accretes material on the main sequence. A 2.45\ms\ star has a main sequence lifetime of $5.65\times10^{8}$\, years, which is shorter than the time it takes the 2.5\ms\ star to reach the end of the AGB. The companion has therefore left the main sequence and is most of the way through the core helium burning phase (with a core helium abundance of around 0.27) before it receives mass from the primary. From this point to the end of helium burning is only another $3.3\times10^{7}$\, years. At this point, the convective envelope is at its shallowest since the end of first dredge-up, and has a thickness of around 0.5\ms. Convection thus dilutes the accreted layer by a factor of about 10. As the star begins to ascend the AGB, the convective envelope again deepens further diluting the accreted material. During this time the maximum mass contained in the envelope is 1.88\ms. 

\begin{table*}
\begin{center}
\begin{tabular}{cccccccccc}
Primary  & Secondary & Case & Dilution & $\log \epsilon$(Li) & \el{12}{C} & \el{12}{C}/\el{13}{C}  & \el{14}{N} & \el{17}{O}/\el{18}{O} & [s/Fe] \\
mass & initial mass & & factor \\
(\ms) & (\ms) \\ 
\hline
2.5 & 2.0 & Std & 0.247 & 0.78 & $ 2.88\times10^{-3} $ & 78.33 & $ 1.40\times10^{-3} $ & 1.58 & 1.14\\ 
& & TH & & -3.27 & $ 2.58\times10^{-3} $ & 14.64 & $ 1.78\times10^{-3} $ & 1.59 & 1.15 \\
& 2.25 &  Std & 0.120  & 1.23 & $ 1.76\times10^{-3} $ & 51.93 & $ 1.24\times10^{-3} $ & 1.74 & 0.87  \\
& & TH & & -0.86 & $ 1.81\times10^{-3} $ & 52.48 & $ 1.24\times10^{-3} $ & 1.76 & 0.89 \\
\hline
3.0 & 2.0 & Std & 0.247 & -0.27 & $ 3.36\times10^{-3} $ & 86.28 & $ 1.70\times10^{-3} $ & 2.10 & 1.19 \\
& & TH & & -6.21 & $ 2.76\times10^{-3} $ & 12.31 & $ 2.43\times10^{-3} $ & 2.10 & 1.19 \\
& 2.25 & Std & 0.126 & -0.13 & $ 2.10\times10^{-3} $ & 59.90 & $ 1.35\times10^{-3} $ & 1.20 & 0.94\\
& & TH & & -5.64 & $ 1.93\times10^{-3} $ & 16.30 & $ 1.61\times10^{-3} $ & 1.22 & 0.95 \\
& 2.4 & Std & 0.050 & 0.03 & $ 1.28\times10^{-3} $ & 37.28 & $ 1.21\times10^{-3} $ & 1.24 & 0.61 \\
& & TH & & -4.05 & $ 1.31\times10^{-3} $ & 36.84 & $ 1.21\times10^{-3} $ & 1.25 & 0.63 \\
& 2.45 & Std & 0.025 & 0.85 & $ 1.02\times10^{-3} $ & 29.87 & $ 1.19\times10^{-3} $ & 1.40 & 0.41 \\
& & TH & & -4.04 & $ 1.03\times10^{-3} $ & 30.54 & $ 1.19\times10^{-3} $ & 1.40 & 0.43 \\
& 2.475 & Std & 0.013 & 0.93 & $ 8.99\times10^{-4} $ & 26.18 & $ 1.18\times10^{-3} $ & 1.28 & 0.25 \\
& & TH & & -3.23 & $ 9.05\times10^{-4} $ & 26.60 & $ 1.18\times10^{-3} $ & 1.29 & 0.27\\
& 2.49 & Std & 0.005 &  0.93 & $ 8.26\times10^{-4} $ & 24.05 & $ 1.17\times10^{-3} $ & 1.22 & 0.12 \\
& & TH & & -2.43 & $ 8.30\times10^{-4} $ & 24.25 & $ 1.17\times10^{-3} $ & 1.22 & 0.14 \\
\hline
4.0 & 2.0 & Std & 0.254 & 1.93 & $ 2.04\times10^{-3} $ & 22.20 & $ 1.47\times10^{-3} $ & 4.28 & 0.95 \\
& & TH & & -3.40 & $ 1.73\times10^{-3} $ & 9.88 & $ 1.84\times10^{-3} $ & 4.25 & 0.96 \\
& 2.25 & Std & 0.121 & 1.98 & $ 1.40\times10^{-3} $ & 22.95 & $ 1.30\times10^{-3} $ & 1.89 & 0.71 \\
& & TH & &  -4.63 & $ 1.31\times10^{-3} $ & 12.66 & $ 1.44\times10^{-3} $ & 1.90 & 0.71 \\
& 2.4 & Std & 0.045 & 2.03 & $ 1.03\times10^{-3} $ & 23.13 & $ 1.20\times10^{-3} $ & 1.14 & 0.43 \\
& & TH & &  -3.45 & $ 1.05\times10^{-3} $ & 21.60 & $ 1.20\times10^{-3} $ & 1.15 & 0.46 \\
& 2.45 & Std & 0.025 & 2.04 & $ 9.00\times10^{-4} $ & 22.94 & $ 1.17\times10^{-3} $ & 1.24 & 0.27 \\
& & TH & & -4.15 & $ 9.12\times10^{-4} $ & 22.85 & $ 1.18\times10^{-3} $ & 1.24 & 0.29 \\
& 2.475 & Std & 0.014 & 1.88 & $ 8.39\times10^{-4} $ & 22.78 & $ 1.17\times10^{-3} $ & 1.22 & 0.16 \\
& & TH & & -3.70 & $ 8.46\times10^{-4} $ & 22.77 & $ 1.17\times10^{-3} $ & 1.22 & 0.18 \\
& 2.49 & Std & 0.005 & 1.56 & $ 8.00\times10^{-4} $ & 22.64 & $ 1.17\times10^{-3} $ & 1.20 & 0.07 \\
& & TH & &  -3.08 & $ 8.03\times10^{-4} $ & 22.64 & $ 1.17\times10^{-3} $ & 1.20 & 0.08 \\
\hline
\end{tabular}
\end{center}
\caption{Surface properties of the models after first dredge-up. No entry is provided for accretion of material from a 2.5\ms\ companion onto stars of between 2.4 and 2.49\ms\ because accretion takes place during core helium burning. `Std' indicates models with convective mixing only; `TH' indicates models where thermohaline mixing is also included. The abundances of \el{12}{C} and \el{14}{N} are given as mass fractions.}
\label{tab:full_details_RGB}
\end{table*}

\begin{table*}
\begin{center}
\begin{tabular}{cccccccccc}
Primary  & Secondary & Case & Dilution & $\log \epsilon$(Li) & \el{12}{C} & \el{12}{C}/\el{13}{C}  & \el{14}{N} & \el{17}{O}/\el{18}{O} & [s/Fe] \\
mass & initial mass & & factor \\
(\ms) & (\ms) \\ 
\hline
2.5 & 2.0 & Std & 0.247 &  0.70 & $ 2.73\times10^{-3} $ & 71.86 & $ 1.49\times10^{-3} $ & 1.74 & 1.13\\
& & TH & & -3.35 & $ 2.46\times10^{-3} $ & 14.33 & $ 1.87\times10^{-3} $ & 1.77 & 1.13 \\
& 2.25 & Std & 0.120 & 1.15 & $ 1.71\times10^{-3} $ & 49.53 & $ 1.31\times10^{-3} $ & 1.92 & 0.86 \\
& & TH & &  -0.95 & $ 1.75\times10^{-3} $ & 50.11 & $ 1.32\times10^{-3} $ & 1.94 & 0.88\\
& 2.4 & Std & 0.052 & 1.28 & $ 1.05\times10^{-3} $ & 35.07 & $ 1.30\times10^{-3} $ & 1.54 & 0.61 \\
& & TH & &  1.21 & $ 1.04\times10^{-3} $ & 34.93 & $ 1.30\times10^{-3} $ & 1.54 & 0.60\\
& 2.45 & Std & 0.026 & 1.12 & $ 9.18\times10^{-4} $ & 28.60 & $ 1.29\times10^{-3} $ & 1.64 & 0.42 \\
& & TH & &  1.11 & $ 9.14\times10^{-4} $ & 28.57 & $ 1.29\times10^{-3} $ & 1.64 & 0.42 \\
& 2.475 & Std & 0.014 & 1.00 & $ 8.36\times10^{-4} $ & 25.27 & $ 1.29\times10^{-3} $ & 1.50 & 0.27 \\
& & TH & &  1.00 & $ 8.35\times10^{-4} $ & 25.29 & $ 1.29\times10^{-3} $ & 1.50 & 0.27\\
& 2.49 & Std & 0.005 & 0.90 & $ 7.77\times10^{-4} $ & 23.20 & $ 1.28\times10^{-3} $ & 1.43 & 0.13 \\
& & TH & &  0.90 & $ 7.77\times10^{-4} $ & 23.20 & $ 1.28\times10^{-3} $ & 1.43 & 0.13\\
\hline
3.0 & 2.0 & Std & 0.247 &  -0.35 & $ 3.15\times10^{-3} $ & 78.15 & $ 1.82\times10^{-3} $ & 2.35 & 1.17 \\
& & TH & & -6.23 & $ 2.60\times10^{-3} $ & 12.06 & $ 2.55\times10^{-3} $ & 2.37 & 1.18 \\
& 2.25 & Std & 0.126 & -0.21 & $ 1.98\times10^{-3} $ & 55.65 & $ 1.46\times10^{-3} $ & 1.37 & 0.92 \\
& & TH & & -5.71 & $ 1.84\times10^{-3} $ & 15.87 & $ 1.71\times10^{-3} $ & 1.40 & 0.94 \\
& 2.4 & Std & 0.050 & -0.06 & $ 1.22\times10^{-3} $ & 35.60 & $ 1.30\times10^{-3} $ & 1.43 & 0.60 \\
& & TH & &  -4.14 & $ 1.25\times10^{-3} $ & 35.43 & $ 1.31\times10^{-3} $ & 1.44 & 0.62\\
& 2.45 & Std & 0.025 & 0.76 & $ 9.75\times10^{-4} $ & 28.71 & $ 1.29\times10^{-3} $ & 1.61 & 0.39 \\
& & TH & &  -4.13 & $ 9.84\times10^{-4} $ & 29.40 & $ 1.29\times10^{-3} $ & 1.62 & 0.42\\
& 2.475 & Std & 0.013 & 0.84 & $ 8.55\times10^{-4} $ & 25.24 & $ 1.28\times10^{-3} $ & 1.49 & 0.24\\
& & TH & &  -3.33 & $ 8.61\times10^{-4} $ & 25.65 & $ 1.28\times10^{-3} $ & 1.49 & 0.26\\
& 2.49 & Std & 0.005 & 0.84 & $ 7.85\times10^{-4} $ & 23.23 & $ 1.28\times10^{-3} $ & 1.42 & 0.12\\
& & TH & & -2.52 & $ 7.88\times10^{-4} $ & 23.42 & $ 1.28\times10^{-3} $ & 1.42 & 0.13\\
\hline
4.0 & 2.0 & Std & 0.254 & 1.84 & $ 1.92\times10^{-3} $ & 21.64 & $ 1.59\times10^{-3} $ & 4.75 & 0.93 \\
& & TH & & -3.49 & $ 1.63\times10^-{3} $ & 9.75 & $ 1.96\times10^{-3} $ & 4.75 & 0.94 \\
& 2.25 & Std & 0.121 & 1.89 & $ 1.33\times10^{-3} $ & 22.27 & $ 1.42\times10^{-3} $ & 2.15 & 0.69 \\
& & TH & & -4.72 & $ 1.24\times10^{-3} $ & 12.42 & $ 1.56\times10^{-3} $ & 2.17 & 0.69 \\
& 2.4 & Std & 0.045 & 1.93 & $ 9.76\times10^{-4} $ & 22.40 & $ 1.31\times10^{-3} $ & 1.32 & 0.42 \\
& & TH & & -3.54 & $ 9.92\times10^{-4} $ & 21.01 & $ 1.31\times10^{-3} $ & 1.33 & 0.44 \\
& 2.45 & Std & 0.025 & 1.94 & $ 8.55\times10^{-4} $ & 22.20 & $ 1.28\times10^{-3} $ & 1.44 & 0.26 \\
& & TH & & -4.25 & $ 8.66\times10^{-4} $ & 22.13 & $ 1.28\times10^{-3} $ & 1.44 & 0.28 \\
& 2.475 & Std & 0.014 & 1.79 & $ 7.97\times10^{-4} $ & 22.04 & $ 1.28\times10^{-3} $ & 1.42 & 0.16\\
& & TH & & -3.79 & $ 8.04\times10^{-4} $ & 22.03 & $ 1.28\times10^{-3} $ & 1.42 & 0.17 \\
& 2.49 & Std & 0.005 & 1.46 & $ 7.60\times10^{-4} $ & 21.90 & $ 1.28\times10^{-3} $ & 1.39 & 0.07 \\
& & TH & & -3.18 & $ 7.63\times10^{-4} $ & 21.90 & $ 1.28\times10^{-3} $ & 1.40 & 0.08 \\
\hline
\end{tabular}
\end{center}
\caption{Surface properties of the models at the start of the TP-AGB. `Std' indicates models with convective mixing only; `TH' indicates models where thermohaline mixing is also included. The abundances of \el{12}{C} and \el{14}{N} are given as mass fractions.}
\label{tab:full_details_AGB}
\end{table*}

Tables~\ref{tab:full_details_RGB} and \ref{tab:full_details_AGB} give the dilution factors at the end of first dredge-up (FDU) and at the start of the thermally pulsing AGB phase for each of our models, as well as the surface abundances of various species and their ratios. Four models begin accreting only once the star has reached the helium-burning phase as described above. These are the 2.4-2.49\ms\ models accreting from the 2.5\ms\ AGB star. None of the 2.49\ms\ stars reaches a final [s/Fe] value that is above the threshold of 0.25 set by \citet{2016MNRAS.459.4299D} as the definition of a barium star (however, the possibility exists for these to be transient barium stars, as discussed below). The 2.475\ms\ star also fails to reach [s/Fe]=0.25 when accreting from a 4\ms\ companion on account of the more massive AGB star producing less s-process enrichment. For the same mass of star, accreting from both the 2.5 and 3\ms\ companions manages to produce a final [s/Fe] value of 0.26, making them barium stars by the barest of margins. The dilution factors remain almost the same when accreting material from different companions because the composition of the material has only a minor influence on the stellar structure.

\subsection{Evolution of surface abundances}

For the majority of isotopes, the evolution of the surface abundances follows that of the dilution factor because most  species are not involved in nuclear reactions at the temperatures found in the envelope regions of these stars. We need therefore only consider the light elements in this discussion. As described by \citet{2007A&A...464L..57S}, in CEMP stars the mixing of carbon into a star's deep interior can lead to FDU enhancing the star's nitrogen content, because the carbon is cycled into this element by the action of proton-capture reactions in the deep interior, and the nitrogen is subsequently dredged to the surface. This result is also true for our Ba stars, though the effects are far less dramatic because the stars already contain a substantial amount of CNO elements. 

\begin{figure}
\includegraphics[width=\columnwidth]{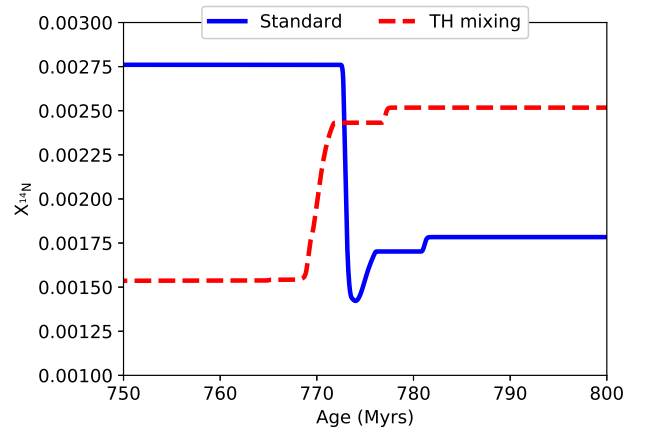}
\caption{Nitrogen-14 abundance as a function of age for the case of accreting 0.5\ms\ from a 3\ms\ companion, around the time of first dredge-up.}
\label{fig:m2M3N14}
\end{figure}

Fig~\ref{fig:m2M3N14} shows the case of accreting 0.5\ms\ from a 3\ms\ companion, and shows the surface abundance\footnote{Unless explicitly state, all abundances are given by mass fraction.} of \el{14}{N} around the time of first dredge-up. With only convection acting, the surface abundance of \el{14}{N} remains at around $2.75\times10^{-3}$ from the end of accretion to the onset of FDU. As the convective envelope deepens, the accreted layer (and the nitrogen along with it) is diluted into the stellar interior, and the nitrogen abundance falls. Eventually, the envelope penetrates those regions of the star that have experience CN cycling, and the nitrogen abundance rises again, reaching a value of around $1.75\times10^{-3}$. When TH mixing is taken into account, the accreted material rapidly becomes diluted into the interior and so the star has a much lower surface nitrogen abundance prior to the onset of FDU. This dilution has also carried a substantial amount of carbon into the star's interior, where it is subject to CN cycling. As FDU happens, the nitrogen abundance therefore rises, reaching a final value of around $2.50\times10^{-3}$. While these changes are not inconsiderable, it is questionable whether they could be detected observationally. The effects are also reduced when less material is accreted, as the mass fraction of carbon reaching the deepest layers is reduced. For accretion masses of 0.1\ms\ or below, material is not transported deep enough by thermohaline mixing to allow the accreted carbon to be processed by proton capture reactions and thus there is no effect on the final surface nitrogen abundance.

In terms of the resulting [N/Fe] values, the models display a narrow range in values from about 0.73 to 1.03. These values are not inconsistent with measurements of nitrogen in barium stars. \citet{2016A&A...586A.151M} report [N/Fe] values of between 0.38 and 1.4 dex, while \citet{1992A&A...262..216B} somewhat lower values from 0.15 to 0.7 dex. Reaching the higher reported values may require the addition of non-convective mixing processes in either the AGB star, or in the barium star itself. In either situation, the circulation of material to hotter regions of the star would allow carbon to be converted to nitrogen via CN-cycling. For one of the stars in their sample -- HD~121447, otherwise known as IT Vir -- \citet{2016A&A...586A.151M} speculate that the high nitrogen abundance is a result of rotationally induced mixing owing to the star being in a close binary system.

The \el{12}{C}/\el{13}{C} isotopic ratio is affected by the inclusion of thermohaline mixing, but the effect is only noticeable when this ratio is high in the accreted material (for accretion from companions of 3\ms\ and below), and a substantial quantity of material is accreted (over 0.25\ms\ or more). Material from the 3\ms\ AGB star has a \el{12}{C}/\el{13}{C} ratio of over 300. When 0.5\ms\ of material is accreted from this star, in the non-TH mixing model the ratio falls to around 80 by the end of FDU. In the TH mixing model, this ratio drops rapidly to around 15 because \el{13}{C} from the interior regions is brought to the surface at the same time as the accreted \el{12}{C} is mixed in. The difference in the post-FDU ratio reduces as the accreted mass reduces. For material from the 4\ms\ AGB star, the \el{12}{C}/\el{13}{C} ratio is already low at around 20. Detecting high \el{12}{C}/\el{13}{C} ratios in Ba stars could potentially discriminate between high and low accretion masses, but only if thermohaline mixing is inefficient or absent. For most of the models we have constructed, a \el{12}{C}/\el{13}{C} ratio in the low to mid-20s is expected which unfortunately means there is little prospect of observationally discriminating between them.

Reported values of the \el{12}{C}/\el{13}{C} ratio in barium stars are typically lower than around 20-30, and often lower than 10 \citep[see e.g.][]{1985ApJ...292..620H, 1992A&A...262..216B, 2018A&A...618A..32K}. It is tempting to suggest that the thermohaline mixing models are preferred because they tend to produce lower \el{12}{C}/\el{13}{C} ratios, but even including therrmohaline mixing does not bring the ratio down low enough to match the lowest observed values. This suggests that something is missing from the AGB models themselves. For the related case of carbon-enhanced metal-poor stars with $s$-process enrichments, \citet{2009MNRAS.396.2313S} point out that AGB models produce \el{12}{C}/\el{13}{C} ratios many orders of magnitude greater than is observed, and they suggest that non-convective mixing on the AGB is likely needed to rectify the discrepancy. The topic of whether such mixing is needed in AGB stars is a matter of some debate. The interested reader is referred to the works of \citet{2010ApJ...713..374K} and \citet{2010ApJ...717L..47B} for a discussion of both sides of the argument. Non-convective mixing that is able to affect the \el{12}{C}/\el{13}{C} ratio will almost certainly alter the lithium abundance (see the discussion below) in the AGB star, and depending on its strength, nitrogen may also be affected too. It is therefore possible that any discrepancies between the light element abundances predicted in the models and their observed counterparts in barium stars is due to uncertain physics in AGB stars. 

\citet{2017A&A...600A..71D} mooted the idea of using the \el{17}{O}/\el{18}{O} isotopic ratio to constrain the initial mass of an AGB star. Provided HBB does not take place, this ratio is determined by the depth of FDU and it changes very little during the course of the AGB. For barium stars, the \el{17}{O}/\el{18}{O} ratio is much more complicated, and depends on the mass accreted and the mass of the companion from which it was accreted. Whether the accreted material undergoes thermohaline mixing does not seem to affect the post-FDU \el{17}{O}/\el{18}{O} ratio. For the 2.5- and 3\ms\ accreta, we find \el{17}{O}/\el{18}{O} ratios between 1.5 and 2.2 and for the 4\ms\ accrete this range is wider, being 1.3 to 4.4. In both cases, larger values occur when more mass has been accreted. Owing to these complications, it is unlike that the \el{17}{O}/\el{18}{O} ratio could be used to obtain any useful information about the mass of the AGB star from which material came. There is very little data on oxygen isotopic ratios in barium stars. A single study by \citet{1985ApJ...292..620H} reports on the determination of \el{16}{O}/\el{17}{O} and \el{16}{O}/\el{18}{O} in six stars. Only in three of these stars (HD~101013, HD~121447 and HD~178717 -- sadly none of which are included in the de Castro sample discussed below) is a value derived for both quantities. Computing the \el{17}{O}/\el{18}{O} ratio for these objects yields values in the range 0.15-3.2. Only the highest accreted masses from the 4\ms\ AGB star are not consistent with this range.

The fragile element lithium may be useful as a way of discriminating between the various models present here. Lithium is easily destroyed at temperatures below around $3\times10^6$\,K which makes it very sensitive to the physics of mixing. Li is most abundant in the ejecta of the 4\ms\ model because it can be produced by the Cameron-Fowler mechanism \citep[see][for a detailed discussion]{2007PASA...24..103K}, while it is moderately depleted in the 3\ms\ model. As described by \citet{2009MNRAS.394.1051S} for the case of carbon-enhanced metal-poor stars, lithium is efficiently destroyed by thermohaline mixing when accretion takes place on the main sequence. Thermohaline convection connects the cool outermost layers of the star with regions where the temperature exceeds the Li destruction temperature. If thermohaline mixing is efficient, we should not expect to detect Li in barium giants at all. However, if thermohaline mixing is not efficient, or accretion takes place during core helium burning, Li may be present at detectable levels. In the former case, Li remaining at the surface will only be diluted into (not destroyed within) the star's convective envelope as it ascends the giant branch. For accretion from the 4\ms\ companion, Li may be present at the level of $\log \epsilon$(Li)\,$\approx1.5-1.9$. For accretion from the 3\ms\ ejecta, $\log \epsilon$(Li) does not exceed 0.9. The case of accretion during core helium burning for the 2.5\ms\ ejecta means that some Li may survive in this case, regardless of the occurrence of TH mixing. Because accretion takes place after the star has developed a cool convective envelope, accreted Li may survive and only become diluted in the envelope, rather than efficiently destroyed. Thus Li at the level of $\log \epsilon$(Li)$>1$ may be indicative of accretion from a lower mass AGB star while the secondary is in the core helium burning phase. However, we caution that Li is a problematic element, whose production and destruction is often very sensitive to the physics involved \citep[e.g.][]{2008ApJ...677..556T, 2010MNRAS.403..505S} and so these results should be treated with some caution. Further issues with Li arise when comparing to abundances determinations from older observational studies. \citet{2002A&A...395L..35R} identified a Ce II line at 6708.099\,\AA\ that was previously unrecognised and was mistakenly identified with Li in observations of post-AGB stars.

\section{Fits to known Ba stars}

We now compare the various models to the properties of known barium stars. We take the catalogue of data from \citet{2016MNRAS.459.4299D}, who assembled a list of 182 objects with their surface properties and abundances. From this list, we restrict our attention to those stars with a similar metallicity to our models. We include an object for comparison if either [Fe I/H] or [Fe II/H] is within 0.125 dex of our fiducial metallicity of [Fe/H]=-0.25. This gives us a list of 74 stars. We then perform a $\chi^2$ fit for each time point in each of our models, considering the effective temperature, $\log g$, and [X/Fe] for the species Na, Mg, Al, Y, Zr, La, Ce, Nd (i.e. only those species likely to be affected by AGB nucleosynthesis). As \citet{2016MNRAS.459.4299D} do not provide errors for the measurements of individual objects, we assume an error of 0.025 dex on $\log \mathrm{T_{eff}}$, 0.25 on  $\log g$ and 0.15 on each [X/Fe]. We caution the reader that a $\chi^2$ fit is a somewhat naive approach, because the various neutron capture species are not truly independent variables.

For each star, we have examined the three best fitting models according to their $\chi^2$ values by eye, partly as a check that the fitting procedure is working as intended, and partly to see how significantly different the best fitting models are. In most cases, there is only a minimal difference in $\chi^2$ between the two best fitting models and these models are the non-thermohaline and thermohaline mixing models for the same accreted mass and mass of the AGB star. This result is not surprising: as described above, there is next to no difference in surface abundances between these cases for the species considered in the fitting process. The data do not support a preference for either the non-thermohaline or thermohaline mixing cases. We have, however, retained this information in the summary table for completeness.

\begin{figure}
\includegraphics[width=\columnwidth]{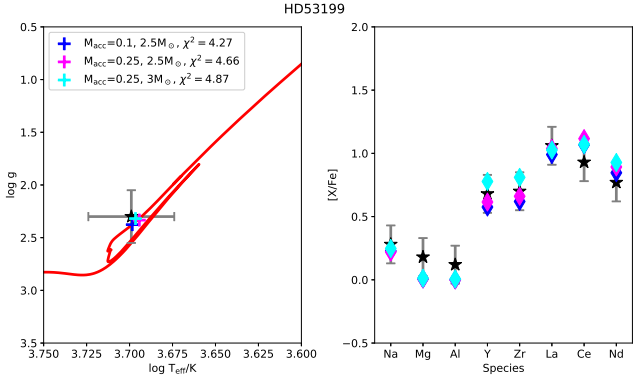}
\caption{Kiel diagram and abundance fits for the star HD~53199, showing the three best fitting models and their associated $\chi^2$ values.}
\label{fig:HD53199}
\end{figure}

\begin{figure}
\includegraphics[width=\columnwidth]{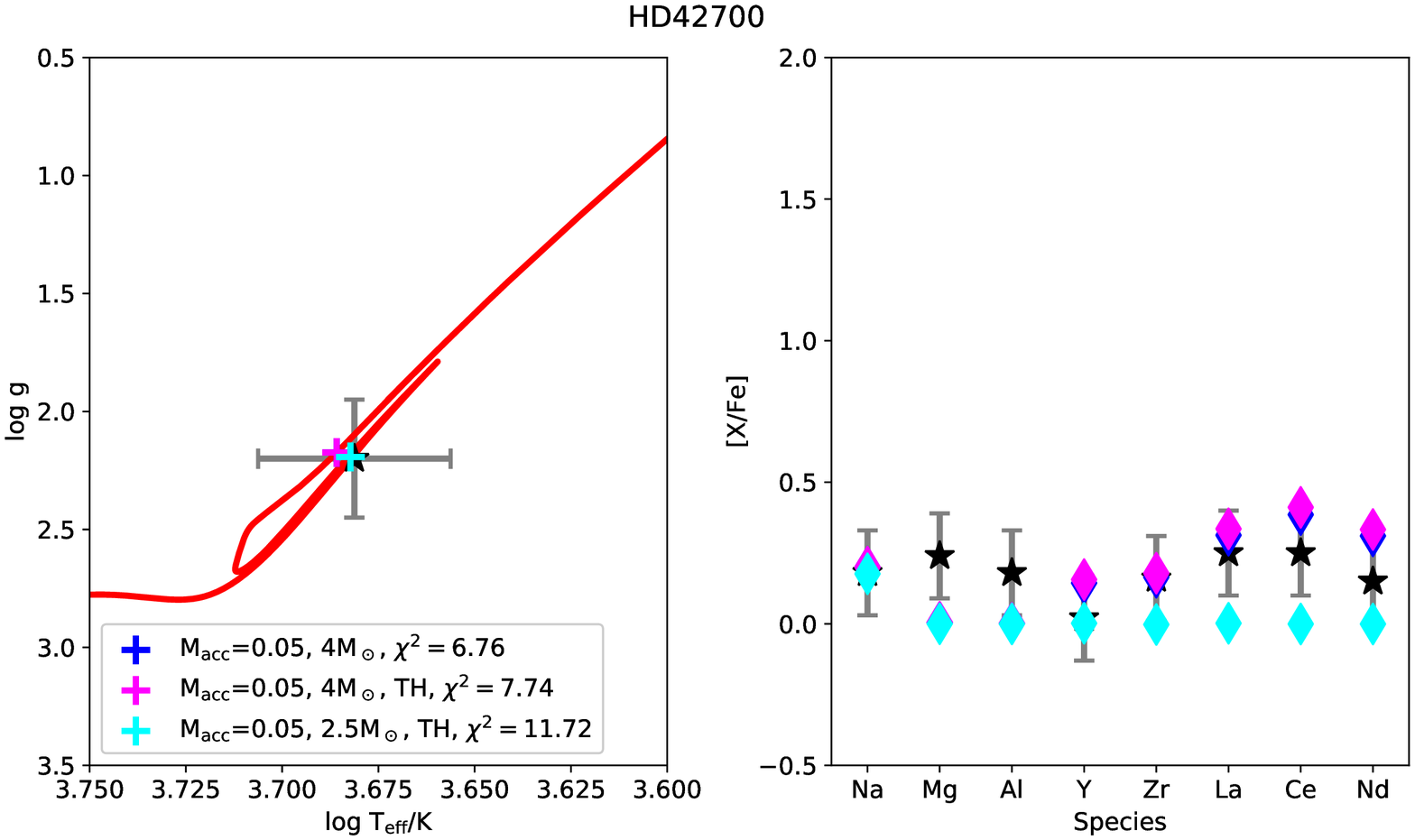}
\caption{Kiel diagram and abundance fits for the star HD~42700, showing the three best fitting models and their associated $\chi^2$ values.}
\label{fig:HD42700}
\end{figure}

\begin{figure}
\includegraphics[width=\columnwidth]{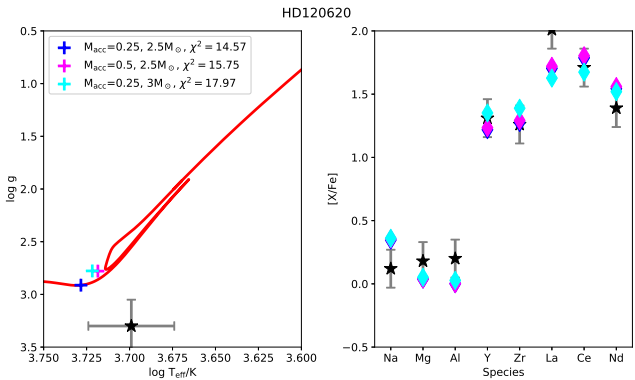}
\caption{Kiel diagram and abundance fits for the star HD~120620, showing the three best fitting models and their associated $\chi^2$ values.}
\label{fig:HD120620}
\end{figure}

Before discussing the sample as a whole, we will describe some individual objects in order to highlight the various successes and failures of the fitting process. 
\begin{itemize}
\item {\it HD~53199:} An excellent fit to the data is obtained for this object, as shown in Fig.~\ref{fig:HD53199}. The object lies toward the end of core helium burning. Accretion of a modest amount of matter from a 2.5\ms\ companion is favoured, and the abundances are all well reproduced.
\item {\it HD~42700:} This object is representative of several stars which display only moderate enhancements of heavy elements. It is not surprising that we find best fits for low accreted mass (higher dilution in the star's envelope), and for material from more massive companions (less s process production). Indeed, the third best fit shown in Fig.~\ref{fig:HD42700} has essentially Solar scaled abundances (with the exception of sodium) and would not be identified as a barium star. Other similar objects include: HD~5322, HD~33709, HD~49778 and HD~95345.
\item {\it HD~120620:} Fig~\ref{fig:HD120620} shows the three best fitting models for this object. While the abundances are tolerably fit (with the exception of lanthanum), the evolutionary track clearly misses the object's $\log g$ and $\log \mathrm{T_{eff}}$. This object may have a lower mass than the 2.5\ms\ we have assumed for our models. \citet{2016MNRAS.459.4299D} give a mass of 2.0\ms\ for this object.
\item {\it HD196445:} This object lies almost exactly on the evolution track for a 2.5\ms\ model. However, a very poor fit to the models is obtained on account of two of the abundances: Mg and La. The object has [Mg/Fe]=0.58 which might point to a more massive AGB star than we have considered, one in which the \el{22}{Ne} source is active and/or HBB is active. However, more massive AGB stars tend to produce less s-process elements. The object also has [La/Fe] = 2.02, which is nearly 1 dex higher than [Ce/Fe] and [Nd/Fe]. With the exception of La, the s-process elements are tolerably well fit by the model.
\end{itemize}

A small group of objects (17 in total) appear to be best fit by very low accretion masses from a 2.5\ms\ companion. However, we should be cautious of this result, as it rests on a na\"ive application of the $\chi^2$ fitting of the models to the data. As described in Section~\ref{sec:models}, when the primary and secondary have similar masses, the secondary is already substantially evolved when it accretes material from the primary AGB star. For secondaries of 2.45\ms\ and above, accretion from the companion 2.5\ms\ primary takes place towards the end of core helium burning -- the phase where many barium giants are found. There is then a brief period of time, lasting less than a few Myrs for the standard model, and even less for the TH mixing model, when the accreted material is not yet fully mixed into the convective envelope. We thus find a good fit to models that have substantial heavy element enrichments, even when accreting small quantities of matter. Given the timescales involved, it seems unlikely we would observe a Ba giant during or immediately after accretion (the presence of the unstable element Tc might confirm this\footnote{\el{99}{Tc} is produced by the $s$ process, and has a half-life of around $10^5$ years. Its presence implies it has been produced recently and is often used to discriminate between intrinsic and extrinsic s-process rich stars \citep[e.g.][]{2020A&A...635L...6S}}, though this would be very difficult to determine as Tc would be mostly ionized in a warm barium star, and Tc {\sc II} lines lie in the far UV; \citealt{1987AJ.....93.1539L}). Many of these fits can also be reproduced to a similar level of confidence with models in which more mass has been accreted.

\begin{table*}
\begin{center}
\begin{tabular}{ccccccccc}
 & & \multicolumn{6}{c}{Companion Mass (\ms)} & Total \\
 & & \multicolumn{2}{c}{2.5} & \multicolumn{2}{c}{3} & \multicolumn{2}{c}{4} \\ 
 & & No TH & TH & No TH & TH & No TH & TH \\
 \hline
 & 0.01 & 9 & 5 & 0 & 0 & 0 & 1 & 15\\
 & 0.025 & 1 & 0 & 0 & 0 & 1 & 0 & 2 \\
 Accreted & 0.05 & 0 & 1 &  2 & 4 & 0 & 1 & 8 \\
 mass & 0.10 & 1 & 0 &  10 & 6  & 1 & 0 & 18 \\
  (\ms) & 0.25 & 6 & 1 & 5 & 6  & 0 & 1 & 19 \\
 & 0.50 & 5 & 3 & 2 & 1 & 1 & 0 & 12 \\
 \hline
 & & 22 & 10 & 19 & 17 & 3 & 3 & \\
 & & \multicolumn{2}{c}{32} & \multicolumn{2}{c}{36} & \multicolumn{2}{c}{6} & 74 \\
 \hline
\end{tabular}
\end{center}
\caption{The number of objects that are best fit by a particular model.}
\label{tab:fitstats}
\end{table*}

Table~\ref{tab:fitstats} shows the number of stars in the sample that are best fit by a particular model, and full details of the fits for each individual system can be found in Tables~\ref{tab:fulldetails1} and \ref{tab:fulldetails2} in the Appendix. With the exception of the cases discussed in the previous paragraph, most of the systems are best fit by systems accreting 0.1\ms\ or more of material. This is to be expected as accreted material is eventually mixed throughout around 80\% of the star. Smaller quantities of accreted material suffer too much dilution and cannot match the high level of heavy element enrichment seen in many of the stars. It is entirely possible (see the discussion of accretion efficiencies below) that a population of former AGB binary systems exist where only a small quantity of material was transferred. That material would become too diluted for the stars to appear as barium stars today. The models favour AGB companions of either 2.5 or 3\ms\ over 4\ms\ companions. Higher mass AGB stars are thought to produce lower quantities of s process elements. Indeed, the objects that are best fit with accretion from a 4\ms\ companion tend to be those with lower heavy element enrichments (such as HD42700, as discussed above). Our results favour higher accretion masses (0.25\ms\ or more) from the 2.5\ms\ AGB star, while for the 3\ms\ companion more modest accretion masses of around 0.1\ms\ are needed.
 
 Our restricted choice of both metallicity and final mass of the Ba star may affect our determinations of the accreted mass. For the metallicity, if the star has [Fe/H] higher than the reference value of -0.25, the model is deficient in iron compared to the object and the computed [X/Fe] value is too high (assuming we have the correct mass fractions for X -- which should be dominated by the AGB ejecta). This will favour a lower accretion mass. As the metallicity range has been restricted to 0.25 dex, this should represent at most a factor of 2 difference between the predicted and actual accreted masses. The order of magnitude of mass accreted should therefore be sound. If the mass of the barium star is higher than our assumed mass of 2.5\ms\ then any accreted material would be diluted in a larger fraction of the star, and a higher accretion mass would be needed to achieve the same [X/Fe].
 
The composition of the ejecta of AGB stars is affected by the choices used to model the complex nucleosynthesis that takes place during the thermally-pulsing phase. The most problematic aspect of AGB nucleosynthesis is how the necessary pocket of \el{13}{C} is formed, as this determines the production of heavy elements. This pocket has been obtained in stellar models in various ways, including adding in convective overshooting at the base of the convective envelope \citep{2000A&A...360..952H,2009ApJ...696..797C}, and including mixing by non-convective phenomena like internal gravity waves \citep{2003MNRAS.340..722D,2016ApJ...827...30B} or magnetic fields \citep{2016ApJ...818..125T}. In the models of \citet{2016ApJ...825...26K} a pocket is produced in the post-processing of the models by adding in a so-called partially mixed zone below the convective envelope at the point that the envelope reaches its maximum depth. The width of this region is a free parameter, which has been set to $2\times10^{-3}$\ms\ for the 2.5 and 3\ms\ models used here, and is half this value for the 4\ms\ model (on account of this star having a narrower intershell region). Within the partially mixed zone, the hydrogen mass fraction declines logarithmically from the envelope value down to $10^{-4}$ at the base of the PMZ. The exact shape of this region affects the production of s elements \citep{2017MNRAS.471..824B}, and larger pockets produce more s elements. In the context of our models, a higher production of s elements in the AGB star means less material needs to be accreted to reach a given level of enrichment in a barium star.
 
Of the objects we have looked at in this study, \citet{2019A&A...626A.127J} provide orbital properties (including masses of the barium star, white dwarf companion and orbital period) for 16 of them. For most of these, our assumed mass for the barium star lies within the uncertainties of their derived values, and the remnant masses are not inconsistent with the AGB companions we have used. However, there are 5 systems that merit further discussion. Two of these systems, HD~88562 and HD~223617 have very low white dwarf masses, of around 0.51\ms. As \citet{2019A&A...626A.127J} note, such low masses are problematic as the primary star needs to be able to reach the AGB and produce s process elements and stellar models do not give third dredge-up at such low core masses. Two systems, HD~49641 and HD~92626, have very large WD masses which would suggest more massive AGB companions than we have looked at. Finally, the system HD~24035 may not fit the wind mass transfer scenario for barium star formation. It has a period of 377.8 days \citep{2019A&A...626A.127J}, which implies a present day separation of around 300\,$\mathrm{R_\odot}$. It is doubtful whether an AGB star could fit within such a tight orbit, and this system may have undergone tidal interactions since the barium star was formed \citep{2020arXiv200505391E}.
 
 \section{Literature comparison}
 
Up to this point, we have only discussed the properties of our objects with respect to the values derived by \citet{2016MNRAS.459.4299D}. \citet{2018A&A...620A.146C} note that several stars in the De Castro sample have high equivalent widths for their La lines. Such high La abundances relative to Ce, the element with the next highest proton number cannot be explained by the s process, which should produce similar quantities of elements around each s process peak. As such, \citet{2018A&A...620A.146C} ignored La in their analysis of this data set. For the 74 stars in the metallicity range we consider, 25 of them have [La/Fe] higher than [C/Fe] by more than 0.45 ($3\sigma$ for our fiducial errorbar). In two stars, HD~24035 and HD~105902, the difference is over 1 dex. Re-running our $\chi^2$ analysis excluding La does not substantially affect our conclusions. Given the problems with the La abundances, it is legitimate to ask how well-determined these abundances of barium stars are. Searching the literature, we have found a handful of objects that have also been studied in detailed by other authors.  The stars we were able to obtain alternative analyses for are: HD~49641, HD~88035, HD~116869, HD~120620,  HD~32712, HD~36653 and HD~211173.

In addition to the study of \citet{2016MNRAS.459.4299D}, HD~49641 has also been studied by \citet{2016RAA....16...19Y} and  \citet{2016MNRAS.463.1213M}. The properties derived by the three groups are listed in Table~\ref{tab:HD49641}. There is discrepancy between the surface gravities report by the three groups, which would put the object in vastly different evolutionary states. \citet{2016RAA....16...19Y} give $\log g = 1.08$, \citet{2016MNRAS.459.4299D} give $\log g = 1.5$ and \citet{2016MNRAS.463.1213M} give $\log g = 3.40$. The former two groups give excellent agreement for [Fe/H], while \citet{2016MNRAS.463.1213M} derive an almost Solar metallicity. If this latter value is correct, comparison with our [Fe/H] = -0.25 models would be inappropriate. Performing a $\chi^2$ fit to the surface properties and abundances provided by \citet{2016RAA....16...19Y} gives a best-fitting model for accretion from a 2.5\ms\ companion, accreting 0.25\ms\ of material, which is the same companion mass as we find with the \citet{2016MNRAS.459.4299D} data. However, the accretion mass for the \citet{2016MNRAS.459.4299D} data requires a higher accretion mass on account of the higher s-process enrichment.
 
 \begin{table}
 \begin{center}
 \begin{tabular}{lccc}
 \hline
 & De Castro & Yang & Mahanta \\
 \hline
 T$_\mathrm{eff}$\,(K) & 4400 & 4645/4351$^a$ & 4700 \\
  $\log g$ & 1.5 & 1.05 & 3.40 \\
  \,[Fe I/H] & -0.30 & - & -0.05 \\
  \,[Fe II/H] & -0.30 & - & -0.02 \\
  \,[Fe/H] & - & -0.32 & \\
  \hline
  [X/Fe] \\
  \hline
  O & - & 0.40 & - \\
  Na &  0.24 & -0.03 & -0.19 \\
  Mg & 0.10 & 0.49 & -0.41 \\
  Al & 0.07 & 0.16 & - \\
  Sr & - & - & 0.97 \\
  Y & 0.89 & 0.35 & 1.31 \\
  Zr & 0.53 & 0.41 & 0.73 \\
  Ba & - & 1.13 & 1.16 \\
  La & 1.86 & 1.38 & 1.50 \\
  Ce & 1.04 & - & 1.66 \\
  Pr & - & - & 1.70 \\
  Nd & 1.14 & - & 1.60 \\
  Eu & - & 0.64 & 0.90 \\
 \hline
 \end{tabular}
 \end{center}
 \caption{Select properties of the star HD~49641, according to the three listed sources. We have listed only surface properties and abundances relevant to AGB nucleosynthesis. $^a$ These temperatures are derived from V-K and B-V respectively.}
 \label{tab:HD49641}
 \end{table}

 The object HD~88035 has also been studied in detail by \citet{2018MNRAS.476.3086K}. These authors determine $T_\mathrm{eff} = 5300$\,K, $\log g = 3.9$, [Fe I/H]  = -0.05 and [Fe II/H] = 0.01. As with HD~49461, these parameters place the object at a hotter, much less evolved stage (cf. $T_\mathrm{eff} = 4900$\,K, $\log g = 2.4$  found by \citet{2016MNRAS.459.4299D}). The higher metallicity found by \citet{2018MNRAS.476.3086K} means our models are not suitable for comparison. In a separate study, \citet{2018A&A...618A..32K} also determined the abundances of two more stars that are found in the \citet{2016MNRAS.459.4299D} sample, and that we have used in this study. These stars are HD~116869 and HD~120620. For the former object, \citet{2018A&A...618A..32K} give the metallicity as [Fe/H]$= -0.44\pm0.12$, placing it outside the metallicity range considered in this study. However, the same authors give the metallicity of HD~120620 as [Fe/H]$=-0.29\pm0.12$ so a comparison with our models can be made. In Table~\ref{tab:HD120620} we summarise the properties of this object. There is general agreement between the surface properties of the two objects, with the more precise measurement of the temperature by \citet{2018A&A...618A..32K} lying within the assumed 300\,K error of the \citet{2016MNRAS.459.4299D} value. The surface gravity values also agree within the uncertainty, while [Fe/H] differs by just 0.01 dex between the two studies. For the elements listed in Table~\ref{tab:HD120620}, the two studies agree to within the uncertainties\footnote{We remind the reader that we have adopted an uncertainty of 0.15 dex for all abundance measurements in the \citet{2016MNRAS.459.4299D} sample, as described earlier.} for all elements with the exception of La and Ce. For cerium, the measurements only overlap at the extremes of the errorbars, with the de Castro value being higher by 0.19 dex. For lanthanum there is a clear discrepancy between the two studies, with \citet{2016MNRAS.459.4299D} determining a value that is 0.41 dex higher than the \citet{2018A&A...618A..32K} value, a gap that cannot be bridged even when considering the uncertainties in both studies. In fact, a lower La abundance is preferable, as Fig.~\ref{fig:HD120620} shows our model struggled to fit the high value reported by \citet{2016MNRAS.459.4299D}. Using the data from the \citet{2018A&A...618A..32K} study, we have re-run our fitting algorithm for HD~120620 using: a) the same subset of elements given in the \citet{2016MNRAS.459.4299D} sample, and b) the full list of heavy elements. In both of these cases we obtain the same results: this object is best fit by accretion from a 2.5\ms\ companion, with an accretion mass of 0.25\ms. This is the same best-fitting model we find when using the \citet{2016MNRAS.459.4299D} data, suggesting that the results of the fitting process are robust when there is general agreement on surface properties and abundances between observational studies. In their study, \citet{2018A&A...618A..32K} found good agreement between the abundance pattern of HD~120620 and either a 2- or 3-\ms\ AGB companion, based on models computed with the {\sc starevol} code. Our finding is therefore in agreement with this earlier study.
 
  \begin{table}
 \begin{center}
 \begin{tabular}{lcc}
 \hline
 & De Castro & Karinkuhzi \\
 \hline
 T$_\mathrm{eff}$\,(K) & 5000 &  $4831\pm13$ \\
  $\log g$ & 3.3 & $3.03\pm0.3$ \\
  \,[Fe I/H] & -0.30 & - \\
  \,[Fe II/H] & -0.30 & -  \\
  \,[Fe/H] & - & $-0.29\pm0.12$ \\
  \hline
  [X/Fe] \\
  \hline
  O & - & $0.51\pm0.17$ \\
  Na &  0.12 & $0.28\pm0.15$ \\
  Mg & 0.18 & $0.30\pm0.24$  \\
  Al & 0.07 & -  \\
  Sr & - & $1.31\pm0.13$  \\
  Y & 1.31 & $1.21\pm0.09$   \\
  Zr & 1.26 & $1.52\pm0.24$  \\
  Ba & - & $2.02\pm0.10$  \\
  La & 2.01 & $1.60\pm0.13$  \\
  Ce & 1.71 & $1.52\pm0.11$  \\
  Pr & - & $1.51\pm0.10$ \\
  Nd & 1.54 & $1.48\pm0.10$  \\
  Eu & - & $0.75\pm0.13$  \\
 \hline
 \end{tabular}
 \end{center}
 \caption{Select properties of the star HD~120620, according to the three listed sources. We have listed only surface properties and abundances relevant to AGB nucleosynthesis. For the Karinkuzhi data, where elements have been measured in more than one ionisation state, the average abundance has been taken and the error chosen to include both measured values.}
 \label{tab:HD120620}
 \end{table}

 \citet{2020MNRAS.492.3708S} made a detailed abundance analysis of three objects which we have used in this study:  HD~32712, HD~36650 and HD~211173. For HD~32712, the surface properties are in good agreement, as are the abundances of the light s-process elements. \citet{2020MNRAS.492.3708S} find an [La/Fe] value of 1.25, compared to 1.56 from \citet{2016MNRAS.459.4299D}. For Ce and Nd, the discrepancy is larger, with  \citet{2020MNRAS.492.3708S} finding higher values by about 0.5 dex. For HD~36650, the two studies have an 0.25 dex discrepancy in [Fe/H], with \citet{2020MNRAS.492.3708S} finding an almost Solar metallicity, placing it outside the range of our study. For HD~211173, the two studies find similar values for $T_\mathrm{eff}$ and $\log g$, but there is a difference of about 0.2 dex in the reported [Fe/H] values (-0.17 versus -0.37). The s process elemental abundances agree to around 0.2 dex with the exception of [La/Fe] which is about 0.4 dex higher in the \citet{2020MNRAS.492.3708S} study. \citet{2020MNRAS.492.3708S} also computed best fits to AGB yields from \citet{2009ApJ...696..797C}, finding best fit models of between 2-3\ms. They also provide dilution factors, which equate to accretion masses of around 0.5\ms\ if the barium star is assumed to have an envelope mass of 2\ms, which is appropriate for 2.5\ms\ star. While we find best fits to similar masses of AGB companion, our accretion masses tend to be lower by a factor of a few because the \citet{2016MNRAS.459.4299D} abundances tend to be lower than those of \citet{2020MNRAS.492.3708S}.
  
\section{Discussion}

A major uncertainty in the study of systems like barium stars and carbon-enhanced, metal-poor stars is the efficiency of mass transfer by stellar winds. Population synthesis models have typically relied upon the Bondi-Hoyle-Lyttleton formula \citep{1944MNRAS.104..273B} to calculate the amount of material a secondary accretes from its primary AGB companion \citep[e.g.][]{1988A&A...205..155B, 2009A&A...508.1359I, 2012A&A...547A..76P}. However, this prescription assumes that the wind speed is fast compared to the binary's orbital speed, a situation unlikely to be satisfied on account of the slow velocities of winds from AGB stars \citep[e.g.][]{1993ApJ...413..641V}.  Detailed hydrodynamical modelling of wind mass transfer systems by \citet{2007ASPC..372..397M} suggests that for slow-moving winds mass transfer may proceed via wind Roche lobe overflow. In this mode, the slow moving wind in a close binary is funnelled through the L1 point of the binary, and may result in much more efficient accretion \citep{2013A&A...552A..26A}. The details of this process are still unclear, with current simulations giving mass accretion efficiencies ranging from 0.01-50\% \citep{2017ApJ...846..117L,2018A&A...618A..50S,2019A&A...626A..68S,2020ApJ...892..110C} depending on mass ratio and separation. \citet{2019A&A...629A.103S} report average mass transfer efficiencies of around 0.01-0.1\% for their simulations of eccentric binaries.

An AGB star of 2.5\ms\ leaves behind a remnant of 0.668\ms\, ejecting a total mass of 1.832\ms\ \citep{2016ApJ...825...26K}.  For the 3\ms\ star, these values are 0.698\ms\ and 2.302\ms\ respectively. For the secondary to accrete 0.5\ms\ of material requires an accretion efficiency of over 20\% from either mass of AGB star. Such high efficiencies are not ruled out by the current hydrodynamical models, but it is questionable whether a substantial number of systems undergo such efficient accretion. However, accretion efficiencies of a few percent would allow the companion to accrete around 0.1\ms\ of material which may be sufficient to produce the observed abundances.

Barium stars have been shown to have a curious property with regard to the masses of the two components of the system. The quantity:
\begin{equation}
Q = \frac{M^3_\mathrm{WD}}{(M_\mathrm{Ba} + M_\mathrm{WD})^2},
\end{equation}
where $M_\mathrm{Ba}$ is the mass of the barium star, and $M_\mathrm{WD}$ is the mass of the white dwarf, has been shown to be strongly peaked for strong barium stars at $Q=0.057\pm0.009$, while mild barium stars exhibit a broader gaussian distribution, with $Q = 0.036\pm0.027$ \citep[see][and references therein for a discussion of this topic]{2019A&A...626A.127J}. In this work, we have fixed the barium star mass at 2.5\ms\, which results in our models having Q values of 0.0297, 0.0333 and 0.0546 for the 2.5-, 3- and 4-\ms\, AGB progenitors respectively. This makes the models coming from our 2.5 and 3\ms\ AGB progenitors incompatible with the Q-values of strong barium stars, while all three progenitors lie within the broader 1-$\sigma$ range for the mild barium stars. This result should be treated with caution as we have fixed the mass of the barium star in this study. A follow-up study with a range of barium star masses (and a range of metallicities) is planned, and we reserve the discussion of Q until the larger model grid is complete.

\begin{figure}
\includegraphics[width=\columnwidth]{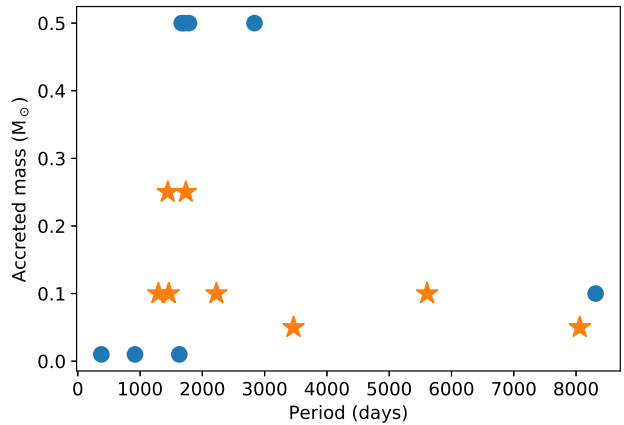}
\caption{Accretion mass as a function of period for the 18 stars in our sample that have periods determined by \citet{2019A&A...626A.127J}. Circles represent systems where the AGB companion is best fit by a 2.5\ms\ model, while stars represent a 3\ms\ companion. }
\label{fig:periods}
\end{figure}

Another question relating to the orbits is whether there is any evidence for a trend in accretion mass as a function of period. It would not be unreasonable to expect that the accretion mass falls off as the period increases. The barium star sample of \citet{2019A&A...626A.127J} provides orbital parameters for 18 objects that we have determined fits for. In Fig~\ref{fig:periods}, we plot the periods and accretion masses for these stars. For stars with a 3\ms\ companion, accretion mass appears relatively flat as a function of period although the two highest accretion masses are found at short period. For systems with a 2.5\ms\ companion, the data is very spread out and it is difficult to draw any useful conclusion. This comparison is somewhat  na\"ive as we have not attempted to account for the eccentricities of the systems, nor have we considered that the period may not represent that of the system during/after mass transfer. Tidal interaction of the system may affect both period and eccentricity after the barium star has formed \citep{2020arXiv200505391E}.

The analysis in this work contains many simplifying assumptions that can be improved in the future. For simplicity, we assumed a representative mass and metallicity for the barium stars. This was done to restrict the number of models that had to be produced. However, the run-times of the models are short (around 1hr or so) and AGB yields are available for a range of metallicities. A larger grid of models could therefore be made and tested against the full \citet{2016MNRAS.459.4299D} data set. We reserve this effort for future work. Rather than using detailed models, an alternative approach would be to use binary population synthesis, as was done for the CEMP-s stars by \citet{2015A&A...581A..22A}. Such models also have the advantage that they explicitly treat the wind accretion process, whereas we have simply assumed accretion masses. However, it is clear from such studies that we are still missing something fundamental in our understanding of the way wind mass transfer works \citep[e.g.][]{2018A&A...620A..63A}, such as failing to account for kicks of nascent white dwarfs \citep{2010A&A...523A..10I} or the interaction of the stars with a circumbinary disc \citep{2013A&A...551A..50D}.

\section{Conclusions}

We have simulated the accretion of material from AGB primaries on to lower mass companion stars, to try and reproduced the observed properties of barium giants. Taking [Fe/H] = -0.25, and assuming 2.5\ms\ is the typical mass of a barium star, we have performed a comparison of our models to 74 stars from the sample of \citet{2016MNRAS.459.4299D}. We find AGB star companions of 2.5 or 3\ms\ best reproduce all but six of the models. Of these remaining objects, all are only mildy enriched in neutron capture elements. We typically require higher accretion mass when accreting from the lower mass companion, and the accretion efficiency required is high, though not inconsistent with hydrodynamical models of wind mass transfer.

\section*{Acknowledgements}

I heartily thank the referee, Alain Jorissen, for his constructive and helpful report, which has improved the quality of this manuscript. This work has made use of the University of Hull's HPC facility, Viper. RJS is funded by the STFC through the University of Hull's Consolidated Grant ST/R000840/1.

\section*{Data availability}

The {\sc STARS} code is available for download from \href{https://people.ast.cam.ac.uk/~stars}{https://people.ast.cam.ac.uk/~stars}. The model sequences generated for this paper are available from the corresponding author upon reasonable request.



\bibliographystyle{mnras}
\bibliography{../../NewBib} 

\appendix

\section{Details of model fits}

Tables~\ref{tab:fulldetails1} and \ref{tab:fulldetails2} give the full details of all the model fits to the 76 stars used in this study. For each object, we provide details of the AGB progenitor mass, total mass accreted and whether the model included thermohaline mixing or not, together with the $\chi^2$ value for the fit. These details are provided for the three best fitting models.

\begin{table*}
\begin{center}
\begin{tabular}{lcccccccccccc}
\hline
& \multicolumn{4}{c}{Best fit} & \multicolumn{4}{c}{$2^\mathrm{nd}$ best fit}  & \multicolumn{4}{c}{$3^\mathrm{rd}$ best fit} \\
 & Progenitor & Accreted & TH &  $\chi^2$ &  Progenitor & Accreted & TH & $\chi^2$ & Progenitor & Accreted & TH & $\chi^2$ \\
Object & mass (\ms) & mass (\ms) & mixing? &  & mass (\ms) & mass (\ms) & mixing? & & mass (\ms) & mass (\ms) & mixing? &\\
\hline
BD-094337  & 3 & 0.5 & No & 11.08 & 3 & 0.25 & No & 12.56 & 3 & 0.1 & No & 14.26 \\
CD-272233  & 3 & 0.25 & Yes & 9.23 & 3 & 0.25 & No & 9.35 & 2.5 & 0.25 & Yes & 11.01 \\
CD-422048  & 3 & 0.5 & Yes & 15.29 & 3 & 0.5 & No & 15.33 & 2.5 & 0.5 & Yes & 17.17 \\
CD-538144  & 3 & 0.25 & No & 11.19 & 3 & 0.25 & Yes & 11.3 & 4 & 0.5 & No & 12.69 \\
CD-611941  & 4 & 0.5 & No & 10.17 & 4 & 0.5 & Yes & 10.21 & 2.5 & 0.25 & No & 10.34 \\
HD15589    & 2.5 & 0.25 & No & 32.62 & 4 & 0.5 & No & 33.32 & 4 & 0.25 & No & 33.76 \\
HD20394    & 3 & 0.1 & No & 9.21 & 2.5 & 0.25 & No & 9.23 & 3 & 0.25 & No & 9.33 \\
HD21989    & 3 & 0.1 & Yes & 9.63 & 3 & 0.1 & No & 9.68 & 2.5 & 0.1 & Yes & 11.69 \\
HD22772    & 3 & 0.25 & No & 5.9 & 3 & 0.25 & Yes & 5.97 & 2.5 & 0.25 & Yes & 9.37 \\
HD24035    & 2.5 & 0.01 & Yes & 52.84 & 2.5 & 0.01 & No & 52.84 & 2.5 & 0.025 & Yes & 53.17 \\
HD29370    & 2.5 & 0.01 & No & 19.7 & 2.5 & 0.5 & No & 21.16 & 4 & 0.5 & No & 21.5 \\
HD32712    & 2.5 & 0.01 & No & 14.17 & 4 & 0.5 & No & 14.83 & 4 & 0.5 & Yes & 15.06 \\
HD36650    & 3 & 0.1 & Yes & 9.31 & 3 & 0.1 & No & 9.43 & 2.5 & 0.1 & Yes & 10.73 \\
HD40430    & 3 & 0.1 & No & 4.26 & 3 & 0.1 & Yes & 4.55 & 3 & 0.05 & No & 5.85 \\
HD49641    & 2.5 & 0.5 & No & 25.8 & 2.5 & 0.5 & Yes & 25.93 & 2.5 & 0.01 & No & 31.82 \\
HD59852    & 3 & 0.05 & No & 5.76 & 3 & 0.05 & Yes & 6.39 & 2.5 & 0.025 & Yes & 6.86 \\
HD66291    & 2.5 & 0.25 & No & 16.94 & 2.5 & 0.01 & No & 17.37 & 4 & 0.25 & Yes & 17.47 \\
HD74950    & 2.5 & 0.25 & No & 14.88 & 4 & 0.25 & Yes & 15.18 & 4 & 0.25 & No & 15.65 \\
HD82221    & 3 & 0.25 & No & 13.97 & 3 & 0.25 & Yes & 14.27 & 2.5 & 0.25 & Yes & 18.4 \\
HD84678    & 2.5 & 0.01 & No & 24.18 & 2.5 & 0.01 & Yes & 26.23 & 2.5 & 0.025 & No & 28.21 \\
HD88035    & 2.5 & 0.5 & Yes & 16.37 & 2.5 & 0.5 & No & 16.41 & 2.5 & 0.25 & No & 17.32 \\
HD88562    & 3 & 0.25 & Yes & 11.16 & 4 & 0.5 & Yes & 11.49 & 4 & 0.5 & No & 11.76 \\
HD91979    & 3 & 0.1 & No & 3.86 & 3 & 0.05 & No & 4.64 & 3 & 0.025 & No & 5.3 \\
HD92626    & 2.5 & 0.01 & Yes & 34.58 & 2.5 & 0.01 & No & 34.59 & 2.5 & 0.025 & Yes & 35.22 \\
HD105902   & 2.5 & 0.01 & No & 51.05 & 2.5 & 0.025 & No & 51.95 & 2.5 & 0.05 & No & 52.5 \\
HD107264   & 2.5 & 0.5 & No & 20.73 & 2.5 & 0.5 & Yes & 21.04 & 2.5 & 0.01 & No & 23.67 \\
HD116869   & 2.5 & 0.25 & No & 9.29 & 2.5 & 0.25 & Yes & 9.49 & 2.5 & 0.1 & No & 9.84 \\
HD120620   & 2.5 & 0.25 & No & 14.57 & 2.5 & 0.5 & No & 15.75 & 3 & 0.25 & No & 17.97 \\
HD143899   & 3 & 0.1 & No & 9.11 & 3 & 0.1 & Yes & 9.58 & 3 & 0.05 & No & 10.37 \\
HD148177   & 3 & 0.1 & Yes & 21.75 & 3 & 0.1 & No & 23.2 & 4 & 0.25 & Yes & 23.59 \\
HD154430   & 2.5 & 0.5 & Yes & 30.95 & 2.5 & 0.5 & No & 31.65 & 3 & 0.5 & Yes & 32.18 \\
HD162806   & 2.5 & 0.25 & Yes & 21.66 & 3 & 0.25 & Yes & 21.9 & 2.5 & 0.01 & No & 21.93 \\
HD168560   & 3 & 0.1 & No & 7.81 & 3 & 0.1 & Yes & 8.36 & 2.5 & 0.01 & No & 8.51 \\
HD168791   & 3 & 0.25 & Yes & 26.52 & 3 & 0.25 & No & 26.76 & 2.5 & 0.25 & Yes & 28.53 \\
HD176105   & 3 & 0.05 & Yes & 3.09 & 3 & 0.05 & No & 3.77 & 2.5 & 0.01 & No & 5.25 \\
HD177192   & 3 & 0.1 & No & 15.98 & 3 & 0.1 & Yes & 16.71 & 3 & 0.05 & Yes & 18.15 \\
HD193530   & 2.5 & 0.01 & No & 18.1 & 2.5 & 0.25 & No & 18.5 & 4 & 0.25 & Yes & 19.14 \\
HD196445   & 3 & 0.5 & Yes & 39.98 & 2.5 & 0.5 & Yes & 40.26 & 3 & 0.5 & No & 40.28 \\
HD201657   & 2.5 & 0.01 & No & 23.17 & 2.5 & 0.5 & No & 24.19 & 2.5 & 0.5 & Yes & 25.03 \\
HD201824   & 2.5 & 0.5 & No & 6.97 & 2.5 & 0.5 & Yes & 7.42 & 2.5 & 0.01 & Yes & 7.58 \\
HD207277   & 2.5 & 0.01 & No & 5.43 & 2.5 & 0.25 & No & 6.01 & 2.5 & 0.25 & Yes & 6.51 \\
HD211173   & 3 & 0.05 & Yes & 7.67 & 3 & 0.05 & No & 7.91 & 2.5 & 0.05 & Yes & 10.26 \\
HD214579   & 2.5 & 0.25 & No & 12.04 & 2.5 & 0.25 & Yes & 12.42 & 3 & 0.25 & No & 13.51 \\
HD217143   & 2.5 & 0.01 & No & 21.36 & 2.5 & 0.5 & No & 22.06 & 2.5 & 0.5 & Yes & 22.21 \\
HD217447   & 3 & 0.25 & No & 4.99 & 3 & 0.25 & Yes & 5.29 & 3 & 0.1 & No & 5.8 \\
HD223617   & 3 & 0.1 & Yes & 11.12 & 3 & 0.1 & No & 11.66 & 2.5 & 0.1 & No & 13.35 \\
HD252117   & 2.5 & 0.5 & Yes & 24.61 & 2.5 & 0.5 & No & 24.75 & 3 & 0.5 & Yes & 25.95 \\
HD273845   & 2.5 & 0.5 & Yes & 10.57 & 2.5 & 0.5 & No & 11.13 & 4 & 0.5 & No & 11.56 \\
BD-18821   & 2.5 & 0.025 & No & 8.49 & 2.5 & 0.05 & No & 9.08 & 2.5 & 0.5 & No & 9.57 \\
HD5322     & 2.5 & 0.01 & Yes & 2.71 & 4 & 0.025 & No & 3.11 & 4 & 0.025 & Yes & 3.23 \\
\hline
\end{tabular}
\end{center}
\caption{Details of the best fits for the first 50 stars in this study. For the best, second-best and third-best fits, we provide the mass of the AGB progenitor, the total mass accreted, whether thermohaline mixing is included and the $\chi^2$ value for the fit.}
\label{tab:fulldetails1}
\end{table*}

\begin{table*}
\begin{center}
\begin{tabular}{lcccccccccccc}
\hline
& \multicolumn{4}{c}{Best fit} & \multicolumn{4}{c}{$2^\mathrm{nd}$ best fit}  & \multicolumn{4}{c}{$3^\mathrm{rd}$ best fit} \\
 & Progenitor & Accreted & TH &  $\chi^2$ &  Progenitor & Accreted & TH & $\chi^2$ & Progenitor & Accreted & TH & $\chi^2$ \\
Object & mass (\ms) & mass (\ms) & mixing? &  & mass (\ms) & mass (\ms) & mixing? & & mass (\ms) & mass (\ms) & mixing? &\\
\hline
HD18182    & 3 & 0.05 & No & 11.29 & 3 & 0.025 & Yes & 11.72 & 3 & 0.05 & Yes & 12.04 \\
HD26886    & 3 & 0.1 & Yes & 6.78 & 3 & 0.1 & No & 6.91 & 2.5 & 0.05 & Yes & 7.11 \\
HD33709    & 3 & 0.05 & Yes & 5.86 & 3 & 0.05 & No & 5.95 & 2.5 & 0.025 & No & 7.52 \\
HD39778    & 3 & 0.1 & No & 3.33 & 3 & 0.05 & No & 4.41 & 2.5 & 0.25 & No & 4.64 \\
HD42700    & 2.5 & 0.01 & Yes & 4.49 & 4 & 0.025 & No & 4.52 & 4 & 0.025 & Yes & 4.54 \\
HD45483    & 3 & 0.1 & No & 11.25 & 3 & 0.1 & Yes & 11.65 & 3 & 0.05 & Yes & 14.32 \\
HD49661    & 2.5 & 0.01 & Yes & 0.75 & 3 & 0.025 & Yes & 1.16 & 3 & 0.025 & No & 1.26 \\
HD49778    & 4 & 0.05 & Yes & 8.14 & 4 & 0.05 & No & 8.31 & 2.5 & 0.025 & Yes & 8.62 \\
HD50075    & 4 & 0.25 & No & 5.53 & 4 & 0.25 & Yes & 5.66 & 3 & 0.1 & No & 6.54 \\
HD50843    & 2.5 & 0.05 & Yes & 8.9 & 4 & 0.1 & Yes & 9.08 & 4 & 0.1 & No & 9.09 \\
HD53199    & 2.5 & 0.1 & No & 4.27 & 2.5 & 0.25 & No & 4.66 & 3 & 0.25 & No & 4.87 \\
HD62017    & 2.5 & 0.01 & No & 18.9 & 3 & 0.1 & Yes & 19.11 & 2.5 & 0.1 & No & 19.59 \\
HD95345    & 4 & 0.025 & No & 5.7 & 4 & 0.025 & Yes & 5.79 & 2.5 & 0.01 & Yes & 6.29 \\
HD113195   & 3 & 0.05 & Yes & 8.89 & 3 & 0.05 & No & 9.51 & 3 & 0.1 & No & 10.45 \\
HD139266   & 3 & 0.25 & Yes & 25.3 & 3 & 0.25 & No & 25.95 & 2.5 & 0.5 & No & 26.17 \\
HD148892   & 3 & 0.25 & No & 13.05 & 3 & 0.25 & Yes & 13.9 & 2.5 & 0.1 & No & 13.93 \\
HD184001   & 3 & 0.1 & Yes & 10.53 & 3 & 0.1 & No & 10.55 & 2.5 & 0.05 & Yes & 13.96 \\
HD102762   & 2.5 & 0.5 & Yes & 31.61 & 2.5 & 0.5 & No & 31.82 & 2.5 & 0.01 & No & 34.16 \\
HD200063   & 3 & 0.25 & Yes & 15.71 & 3 & 0.25 & No & 15.72 & 2.5 & 0.25 & Yes & 17.0 \\
HD214889   & 3 & 0.1 & No & 5.16 & 3 & 0.1 & Yes & 5.32 & 3 & 0.05 & No & 6.01 \\
HD749      & 3 & 0.25 & Yes & 16.98 & 3 & 0.25 & No & 17.19 & 3 & 0.1 & No & 19.23 \\
HD211221   & 4 & 0.01 & Yes & 5.4 & 4 & 0.01 & No & 5.44 & 3 & 0.01 & No & 7.03 \\
HD89638    & 3 & 0.1 & No & 3.04 & 3 & 0.1 & Yes & 3.2 & 2.5 & 0.05 & No & 3.28 \\
HD187762   & 4 & 0.1 & No & 9.24 & 2.5 & 0.025 & No & 9.36 & 4 & 0.1 & Yes & 9.43 \\
\hline
\end{tabular}
\end{center}
\caption{Details of the best fits for the final 24 stars in this study. For the best, second-best and third-best fits, we provide the mass of the AGB progenitor, the total mass accreted, whether thermohaline mixing is included and the $\chi^2$ value for the fit.}
\label{tab:fulldetails2}
\end{table*}

\bsp	
\label{lastpage}
\end{document}